\documentclass[review]{elsarticle}
\makeatletter
\def\ps@pprintTitle{%
 \let\@oddhead\@empty
 \let\@evenhead\@empty
 \def\@oddfoot{}%
 \let\@evenfoot\@oddfoot}
\makeatother
\usepackage{amssymb}
    \usepackage{graphicx}
         \usepackage{color}
    \usepackage{caption,subfig}
\usepackage[cmex10]{amsmath}
\usepackage{natbib}

\begin{document}

\begin{frontmatter}

\title{Robust Tomlinson-Harashima Precoding for Two-Way Relaying}

\cortext[cor1]{Corresponding author}
\author{Saeideh Mohammadkhani\corref{cor1}}
\ead{s\_mohammadkhani@elec.iust.ac.ir}
\author{Amir Hosein Jafari\corref{cor2}}
\address{Electrical Engineering Department, Iran University of Science and Technology, Tehran, Iran }
\ead{amirjafari@iust.ac.ir}

\author{George K. Karagiannidis, {Fellow, IEEE}\corref{cor2}}
\address{Department of Electrical and Computer Engineering, Aristotle University of Thessaloniki, Thessaloniki 541 24, Greece}
\ead{geokarag@auth.gr}
\begin{abstract}
Most of the non-linear transceivers, which are based on Tomlinson Harashima
(TH) precoding and have been proposed in the literature for two-way relay networks, assume perfect channel state information (CSI). In this paper, we propose a novel and robust TH precoding scheme for two-way relay networks with multiple antennas at the transceiver and the relay nodes. We assume imperfect CSI and
 that the channel uncertainty is bounded by
a spherical region. Furthermore, we consider the sum of the mean square error as the objective function, under a limited power constraint for transceiver and relay nodes. Simulations are provided to evaluate the
performance and to validate the efficiency of the proposed scheme.
\end{abstract}


\end{frontmatter}


\section{Introduction}
By using network coding, two-way relay networks have attracted a significant attention, due to its advantage in terms of spectral efficiency \cite{kara}. 
On the other hand, multiple$-$input$-$multiple$-$output (MIMO) technique 
 enhances spatial diversity, throughput and reliability. The combination of MIMO and two-way relaying with
 precoding at both source and relay nodes shows the benefits of them. In addition, non-linear precoding at the transmitter, in the form of  TH
 with linear relay precoder and linear minimum mean square error (MMSE) receiver provides a better bit error-rate (BER) performance in comparison to linear source precoder \cite{utschick}.

The performance of a MIMO relaying system depends on the available channel state information (CSI). However, in most practical cases, CSI is imperfect, due to quantization error or inaccurate channel
estimation, which is a result of insufficient training sequences
or low signal-to-noise ratio (SNR), feedback errors, etc.. Therefore, this imperfectness must be explicitly
considered in the estimated channel that is led to some robust
designs, which are less sensitive to the optimization errors. In general, there are two types of robust designs: \textit{stochastic} \cite{saeideh1} and \textit{worst case} \cite{sm}.
In the worst-case, the channel error
is considered to belong to a predefined uncertainty region and
the final goal is the optimization of the worst system performance for each error in this region. In the stochastic approach, a stochastic viewpoint
is chosen to look to the problem and the required robustness is
acquired from a probabilistic feature. Regarding the stochastic
approach advantages, the worst-case design is necessary to
take absolute robustness, i.e., guaranteed performance with
probability one.

TH precoding is more sensitive on the channel estimation errors, compared to linear precoding techniques, due to its nonlinear nature. Specifically, in the presence
of channel imperfectness, the performance of TH precoding would be deteriorated critically \cite{utschick}. In \cite{2}-\cite{4}, robust linear precoding were considered for one way network. TH precoding design in one way relay network with perfect CSI was proposed in \cite{Wu} and robust consideration were done in \cite{Wu1}-\cite{Gopal}. However, to our best knowledge, the research on robust TH precoding design for two way MIMO networks is missing.

In this paper, we propose a TH precoding scheme for two-way MIMO relay systems, where both source and relay nodes are equipped with multiple antennas. Furthermore, perfect CSI of the source-relay links and imperfect CSI  of the relay-destination links are available at the relay node. The aim is to minimize the sum of mean square error (MSE) at each receiver node, keeping the transmit power of relay and source nodes less than a threshold.

\textit{Notations}- The lower case and upper case boldface letters indicate the vectors and matrices, respectively.
     $(.)^H$, $(.)^T$, $(.)^{-1}$, $||.||$, $|.|$, $tr(.)$, $\mathbb{E}(.)$ and $\mathbf{I}_{_N}$ represent Hermitian, Transpose, inversion, Frobenius norm, determinant, trace of a matrix, statistical expectation, and an identity matrix of size $N$, respectively.

\section{System Model And Problem Description}
    \begin{figure}
    \begin{center}
    \includegraphics[width=1\textwidth]{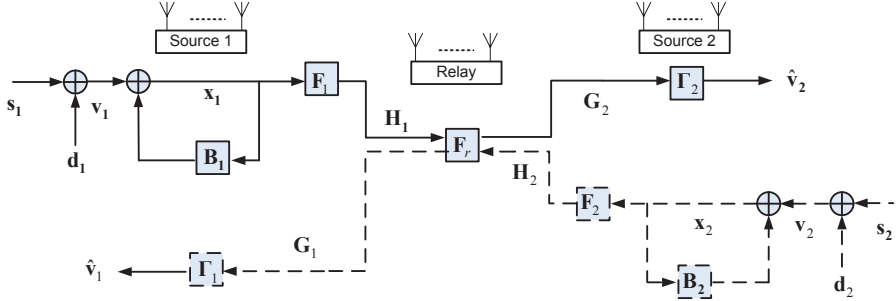}
     \caption{MIMO Two-Way Relay with TH Precoding.}\label{fig1}
     \end{center}
    \end{figure}
    We consider a MIMO two way relay system including two multiple antenna nodes with $N_t$ antennas, which exchange their information with the help of one relay node, equipped with $N_r$ antennas as shown in Fig. 1. The information exchange between nodes 1 and 2 is performed in two time slots. In the first, nodes 1 and 2 concurrently fed their information, $\mathbf{s}_i=[s_{i,1},...s_{i,N_t}]$, into the TH precoder. The resulted vector of signals at each transmitter node is
    \begin{equation}\label{eq1}
    \mathbf{x}_i=\mathbf{C}^{-1}_i \mathbf{v}_i,
    \end{equation}
    where
       $\mathbf{C}_i=\mathbf{B}_i+\mathbf{I}_{N_t}$ is a lower left triangular matrix with unit diagonal elements and $\mathbf{v}_i=\mathbf{s}_i+\mathbf{d}_i$ contains modified data symbols, where $\mathbf{d}_i$ is such that the real and imaginary components of $\mathbf{x}_i$ are constrained to be within the region $(-\sqrt{M},\sqrt{M}]$ which $M$ is the number of constellation points in the M-ary QAM modulation scheme. In addition, the entries of $\mathbf{x}_i$ is considered as $\mathbb{E}(\mathbf{x}_i\mathbf{x}^H_i)={\sigma}^2_{\mathbf{x}_i}\mathbf{I}$. After the nonlinear operation, the vector $\mathbf{x}_i$ is multiplied with an $N_t\times N_t$ precoder matrix $\mathbf{F}_i, i=1,2$ and forward to relay. The received signals at the relay antennas are
    \begin{equation}\label{eq3}
    \mathbf{Y}_r=\textbf{H}_1 \textbf{F}_1 \textbf{x}_1+\textbf{H}_2 \textbf{F}_2 \textbf{x}_2+\textbf{n}_r,
    \end{equation}
     where $\mathbf{H}_i, i=1,2$ is the $N_r \times N_t$ channel matrix between the node $i$ and the relay node and $\mathbf{n}_r$ is the additive white complex Gaussian noise vector at relay with $\sigma^2_{n_r}$.
  In the second time slot, the received signal in the relay is multiplied by an $N_r\times N_r$ linear precoding matrix $\mathbf{F}_r$ and forward to the receivers.
\begin{equation}\label{eq4}
\mathbf{x}_r=\mathbf{F}_r\mathbf{Y}_r=\textbf{F}_r\textbf{H}_1 \textbf{F}_1 \textbf{x}_1+\textbf{F}_r\textbf{H}_2 \textbf{F}_2 \textbf{x}_2+\textbf{F}_r\textbf{n}_r.
\end{equation}
The received signal at the receivers can be written as
\begin{equation}\label{eq5}
\textbf{y}_i=\textbf{G}_i\textbf{F}_r\textbf{H}_1 \textbf{F}_1 \textbf{x}_1+\textbf{G}_i\textbf{F}_r\textbf{H}_2 \textbf{F}_2 \textbf{x}_2+\textbf{G}_i\textbf{F}_r\textbf{n}_r+\mathbf{n}_i,
\end{equation}
where $\mathbf{G}_i,i=1,2$ is $N_t\times N_r$ channel matrix between relay and $i$th receiver.

The following assumptions are made about the CSI:
\begin{enumerate}
\item The receiver nodes have available perfect CSI of the equivalent channels between transmitter-relay-receiver, $\mathbf{G}_i \mathbf{F}_r \mathbf{H}_i\mathbf{F}_i$, $\mathbf{G}_i\mathbf{F}_r\mathbf{H}_{j}\mathbf{F}_{j}$ that $ i,j=1,2 , i\neq j$. The equivalent channel can be estimated by using the sent training sequence from transmitter and received in receiver after passing from relay.
\item	The source-relay channel, $\mathbf{F}_i$, is perfected estimated at the relay by using a training sequence.
\item The relay-receiver CSI is not perfect at the relay, due to limitation in the rate of feedback link from receiver to relay or due to feedback error.
\end{enumerate}
Based on above assumptions, the self-interference can be completely removed. Therefore,
\begin{eqnarray}\label{eq6}
    \bar{\mathbf{Y}}_i=\mathbf{G}_i\mathbf{F}_r\mathbf{H}_j \mathbf{F}_j \mathbf{x}_j+\mathbf{G}_i\mathbf{F}_r\mathbf{n}_r+\mathbf{n}_i, i,j=1,2, i\neq j
      \end{eqnarray}

      Due to its simplicity, a linear receiver is used at each receiver to retrieve the transmitted signals. Denoting $\boldsymbol{\Gamma}_i$ as the
      $N_t\times N_t$ matrix at the $i$ receiver, the estimation of the transmitted signal vector can be expressed as
       \begin{eqnarray}\label{eq7}
             \hat{\mathbf{v}}_i=\boldsymbol{\Gamma}_i\mathbf{G}_1\mathbf{F}_r\mathbf{H}_j \mathbf{F}_j \mathbf{x}_j+\boldsymbol{\Gamma}_i\mathbf{G}_i\mathbf{F}_r\mathbf{n}_r+\boldsymbol{\Gamma}_i\mathbf{n}_i,
      \end{eqnarray}
      where $i,j=1,2,i \neq j$. Note that if $\mathbf{v}_i$ can be estimated at the destination, $\mathbf{s}_i$ can be recovered by modulo operation.

In this paper, we consider the minimization of the sum MSE of two receiver nodes in order to estimate $\mathbf{v}_i$ subject to transmit power constraint at the relay and transmitter nodes. Optimization is jointly done over TH precoding matrices $\mathbf{C}_i$, $\mathbf{F}_i$, linear relay precoder $\mathbf{F}_r$ and linear equalizer at the receiver $\boldsymbol{\Gamma}_i$. Thus the optimization problem can be formulated as
\begin{eqnarray}\label{eq8}
&\underset{\boldsymbol{\Gamma}_i, \mathbf{F}_i, \mathbf{F}_r,\mathbf{C}_i, i=1,2}  {\text{min}}&
  \quad mse_1+mse_2  \nonumber \\
& \text{s.t.}&     P_T\leq P_{r,t}, P_1\leq P_{1,t}, P_2\leq P_{2,t},
\end{eqnarray}
where $mse_i$ is the MSE at the $i$th receiver, $P_T$ , $P_i,i=1,2$ are the transmit power of relay and $i$th transmitter and $P_{r,t}$, $P_{i,t}$ are the maximum power which can be used by the relay and $i$th transmitter.

The MSE at the $i$th receiver node can be written as
    \begin{equation}\label{eq9}
    mse_i=\mathbb{E}(\|\hat{\mathbf{v}}_i-\mathbf{C}_j\mathbf{x}_j\|^2)
=\mathbb{E}(\|(\boldsymbol{\Gamma}_i\mathbf{G}_i\mathbf{F}_r\mathbf{H}_j \mathbf{F}_j
-\mathbf{C}_j)\mathbf{x}_j\|^2)\\
+\sigma^2_{n_r}(\|\boldsymbol{\Gamma}_i\mathbf{G}_i\mathbf{F}_r\|^2)+\sigma^2_{n_i}\|\boldsymbol{\Gamma}_i\|^2,
\end{equation}
where $j=2$ if $i=1$ and $j=1$ if $i=2$.
The transmit power of the relay node is
\begin{eqnarray}\label{eq10}
       P_{T}&=\sigma^2_{x_1}tr(\mathbf{F}_r\mathbf{H}_1\mathbf{F}_1\mathbf{F}^H_1\mathbf{H}^H_1\mathbf{F}^H_r)+\sigma^2_{x_2}tr(\mathbf{F}_r\mathbf{H}_2\mathbf{F}_2\mathbf{F}^H_2\mathbf{H}^H_2\mathbf{F}^H_r)
       +\sigma^2_{n_r}tr(\mathbf{F}_r\mathbf{F}^H_r)\nonumber\\
\end{eqnarray}
The transmit power of $i$th transmitter node can be denoted as
\begin{eqnarray}\label{eq11}
P_i=\sigma^2_{x_i}tr(\mathbf{F}_i\mathbf{F}^H_i), i=1,2.
\end{eqnarray}

We assume that the information for the channels between relay-receivers are not perfect. Therefore, by considering the popular methods for channel estimation,
we have
\begin{eqnarray}\label{eq12}
\mathbf{G}_i=\hat{\mathbf{G}}_i+\mathbf{\Delta} \mathbf{G}_i,
\end{eqnarray}
where $\hat{\mathbf{G}}_i$ is the estimated channels and $\Delta \mathbf{G}_i$ is the channel error matrice that is bounded by spherical, i.e.
   \begin{eqnarray}\label{eq13}
     \mathcal{S}_g=\{\mathbf{a} \in \mathbb{C}:||\mathbf{a}||^2\leq  \sigma_{g_i}^2\},
    \mathbf{\Delta}\boldsymbol{\mathbf{G}} \in \mathcal{S}_g.
     \end{eqnarray}
     It should be noted that the actual error is unknown and only the upper bound, $\varepsilon_g^2$ is known. When the channel error exists, there are infinite goals and constraints for the problem and it is unsolvable. In the rest of paper, we attempt to obtain a solution for (\ref{eq8}) with CSI errors.
     \section{Robust THP Design}
     To solve the optimization problem in (\ref{eq8}), using a worst-case design, we could transform it to a simpler problem as
      \begin{eqnarray}\label{eq14}
& \underset{\boldsymbol{\Gamma}_i, \mathbf{F}_i, \mathbf{F}_r, \mathbf{C}_i, i=1,2}{\text{min}}&
\quad \underset{\mathbf{\Delta}\mathbf{G}_i}{\text{max}}\quad mse_1+mse_2\nonumber \\
& \text{s.t.}& P_T\leq P_{r,t}, P_1\leq P_{1,t}, P_2\leq P_{2,t}.
\end{eqnarray}
It can see from (\ref{eq14}), the optimization is done over the error of the channels. To this end, the channel error is considered in sum MSE expression.
\begin{equation}\label{eq15}
\begin{aligned}[b]
      \underset{\mathbf{\Delta}\mathbf{G}_i}{\text{max}}~~~&mse_i\leq \sigma^2_{x_j}\|(\boldsymbol{\Gamma}_i\hat{\mathbf{G}}_i\mathbf{F}_r\mathbf{H}_j \mathbf{F}_j-\mathbf{C}_j\|^2\\
&+\sigma^2_{x_j}\sigma^2_{g_i}\|\boldsymbol{\Gamma}_i\|^2\|\mathbf{F}_r\mathbf{H}_j \mathbf{F}_j\|^2
      +\sigma^2_{n_r}\|\boldsymbol{\Gamma}_i\hat{\mathbf{G}}_i\mathbf{F}_r\|^2\\
&+\sigma^2_{n_r}\sigma^2_{g_i}\|\boldsymbol{\Gamma}_i\|^2\|\mathbf{F}_r\|^2
      +\sigma^2_{n_i}\|\boldsymbol{\Gamma}_i\|^2.
      \end{aligned}
      \end{equation}
By considering this fact that the power constraints are not related to $\boldsymbol{\Gamma}_i$, we minimize the obtained sum $mse$ over $\boldsymbol{\Gamma}_i$
      \begin{equation}\label{eq16}
      \begin{aligned}[b]
      \frac{\partial}{\partial\boldsymbol{\Gamma}^*_i}=0\Rightarrow &\sigma^2_{x_j}\boldsymbol{\Gamma}_i \hat{\mathbf{G}}_i\mathbf{F}_r\mathbf{H}_j\mathbf{F}_j\mathbf{F}^H_j\mathbf{H}^H_j\mathbf{F}^H_r\hat{\mathbf{G}}^H_i
      -\sigma^2_{x_j}\mathbf{C}_j\mathbf{F}^H_j\mathbf{H}^H_j\mathbf{F}^H_r\hat{\mathbf{G}}^H_i\\
      &+\sigma^2_{x_j}\sigma^2_{g_i}\boldsymbol{\Gamma}_i\|\mathbf{F}_r\mathbf{H}_j\mathbf{F}_j\|^2+\sigma^2_{n_r}\boldsymbol{\Gamma}_i
      \hat{\mathbf{G}}_i\mathbf{F}_r\mathbf{F}^H_r\hat{\mathbf{G}}^H_i
+\sigma^2_{n_r}\sigma^2_{g_i}\boldsymbol{\Gamma}_i\|\mathbf{F}_r\|^2
      +\sigma^2_{n_i}\boldsymbol{\Gamma}_i=0,
      \end{aligned}
      \end{equation}
      and
      \begin{equation}\label{eq17}
      \begin{aligned}[b]
      \boldsymbol{\Gamma}_i=&\sigma^2_{x_j}\mathbf{C}_j\mathbf{F}^H_j\mathbf{H}^H_j\mathbf{F}^H_r\hat{\mathbf{G}}^H_i
      \\&\times(\sigma^2_{x_j}\hat{\mathbf{G}}_i\mathbf{F}_r\mathbf{H}_j\mathbf{F}_j\mathbf{F}^H_j\mathbf{H}^H_j\mathbf{F}^H_r\hat{\mathbf{G}}^H_i
+\sigma^2_{x_j}\sigma^2_{g_i}\|\mathbf{F}_r\mathbf{H}_j\mathbf{F}_j\|^2\mathbf{I}\\
&+\sigma^2_{n_r}
      \hat{\mathbf{G}}_i\mathbf{F}_r\mathbf{F}^H_r\hat{\mathbf{G}}^H_i+\sigma^2_{n_r}\sigma^2_{g_i}\|\mathbf{F}_r\|^2\mathbf{I}
      +\sigma^2_{n_i}\mathbf{I})^{-1}.
     \end{aligned}
      \end{equation}
      By replacing the obtained $\boldsymbol{\Gamma}_i$ in the (\ref{eq15}), we obtain the following expression for $mse$
       \begin{equation}\label{eq18}
      \begin{aligned}
       mse_i=&tr(\sigma^2_{x_j}\mathbf{C}_j(\mathbf{I}-\sigma^2_{x_j}\mathbf{D}^H_j\mathbf{E}^H_i({A}_i\mathbf{I}+\mathbf{B}_i+\mathbf{E}_i\mathbf{D}_j\mathbf{D}^H_j\mathbf{E}^H_i)^{-1}
      \times \mathbf{E}_i\mathbf{D}_j)\mathbf{C}^H_j),
      \end{aligned}
      \end{equation}
      where
      \begin{equation}\label{eq19}
      \begin{aligned}[b]
      A_i&=\sigma^2_{x_j}\sigma^2_{g_i}||\mathbf{F}_r\mathbf{D}_j||^2+\sigma^2_{n_r}\sigma^2_{g_i}||\mathbf{F}_r||^2+\sigma^2_{n_i},\\
      \mathbf{E}_i&=\hat{\mathbf{G}}_i\mathbf{F}_r,\\
      \mathbf{D}_i&=\mathbf{H}_i\mathbf{F}_i,\\
      \mathbf{B}_i&=\sigma^2_{n_r}\mathbf{E}_i\mathbf{E}^H_i.
      \end{aligned}
      \end{equation}
    By applying the matrix inversion lemma $(A+BCD)^{-1}=A^{-1}-A^{-1}B(DA^{-1}B+C^{-1})BA^{-1}$, we obtain
     \begin{equation}\label{eq20}
      mse_i=tr(\sigma^2_{x_j}\mathbf{C}_j(\mathbf{I}+\sigma^2_{x_j}\mathbf{D}^H_j\mathbf{E}^H_i(A_i\mathbf{I}+\mathbf{B}_i)^{-1}\mathbf{E}_i\mathbf{D}_j)^{-1}\mathbf{C}^H_j).\\
      \end{equation}

    In the second step, the optimization must be done over $\mathbf{F}_i,\mathbf{F}_r, \mathbf{C}_i$. Since the problem (10) by considering (\ref{eq20}) is nonconvex, a globally optimal solution of $\mathbf{F}_i,\mathbf{F}_r, \mathbf{C}_i$ is difficult to obtain with reasonable computational complexity. We develop an iterative algorithm. Before doing optimization, using the relation between trace and determinant, the MSE expression is changed. Indeed, a lower bound for MSE is considered as
     \begin{eqnarray}\label{eq21}
     mse_i=|\mathbf{I}+\sigma^2_{x_j}\mathbf{D}^H_j\mathbf{E}^H_i(\mathbf{A}_i\mathbf{I}+\mathbf{B}_i)^{-1}\mathbf{E}_i\mathbf{D}_j|^{\frac{-1}{N}}.
     \end{eqnarray}
     In this relation, we use  $tr(\mathbf{X})\geq N|\mathbf{X}|^{\frac{1}{N}}$ with a $N\times N$ positive semidefinite matrix, $\mathbf{X}$. If the $\mathbf{X}$ is diagonal and have equal diagonal elements, this relation is established for equality. Here, we use this fact $|\mathbf{C}_i\mathbf{C}^H_i|=1$ and $|\mathbf{X}\mathbf{Y}|=|\mathbf{Y}\mathbf{X}|$. In addition, since minimizing $|\mathbf{X}|^{(-1)}$ is equivalent to maximizing $|\mathbf{X}|$, therefore, the equivalent equation can be expressed as
   \begin{equation}\label{eq22}
\begin{aligned}
\underset{\mathbf{F}_1,\mathbf{F}_2,\mathbf{F}_r}{\text{max}} &|\mathbf{I}+\sigma^2_{x_2}\mathbf{D}^H_2\mathbf{E}^H_1(\mathbf{A}_1\mathbf{I}+\mathbf{B}_1)^{-1}\mathbf{E}_1\mathbf{D}_2|
+|\mathbf{I}+\sigma^2_{x_1}\mathbf{D}^H_1\mathbf{E}^H_2(\mathbf{A}_2\mathbf{I}+\mathbf{B}_2)^{-1}\mathbf{E}_2\mathbf{D}_1|\\
\text{s.t.}~~~&
\sigma^2_{x_1}tr(\mathbf{F}_r\mathbf{D}_1\mathbf{D}^H_1\mathbf{F}^H_r)+\sigma^2_{x_2}tr(\mathbf{F}_r\mathbf{D}_2\mathbf{D}^H_2\mathbf{F}^H_r)
       +\sigma^2_{n_r}tr(\mathbf{F}_r\mathbf{F}^H_r)\leq P_{r,t}\\
       &\sigma^2_{x_i}tr(\mathbf{F}_i\mathbf{F}^H_i)\leq P_{i,t}, i=1,2.
      \end{aligned}
      \end{equation}
     In order to solve the equivalent master problem, we consider the following singular value decomposition (SVD)
      \begin{equation}\label{eq22}
      \begin{aligned}[b]
      \mathbf{H}&=[\mathbf{H}_1,\mathbf{H}_2]=\mathbf{U}_h\boldsymbol{\Lambda}^{(1/2)}_h\mathbf{V}^H_h,\\
      \mathbf{H}_1&=\mathbf{U}_h\boldsymbol{\Lambda}^{(1/2)}_{h}\mathbf{V}^H_{h,1},\\
      \mathbf{H}_2&=\mathbf{U}_h\boldsymbol{\Lambda}^{(1/2)}_h\mathbf{V}^H_{h,2},\\
      \mathbf{V}^H_{h}&=[\mathbf{V}^H_{h,1},\mathbf{V}^H_{h,2}],
      \end{aligned}
      \end{equation}
and
      \begin{equation}\label{eq23}
      \begin{aligned}[b]
      \hat{\mathbf{G}}&=[\hat{\mathbf{G}}^T_1,\hat{\mathbf{G}}^T_2]^T=\mathbf{U}_g\boldsymbol{\Lambda}^{(1/2)}_g\mathbf{V}^H_g,\\    \hat{\mathbf{G}}_1&=\mathbf{U}_{g,1}\boldsymbol{\Lambda}^{(1/2)}_g\mathbf{V}^H_{g},\\
      \hat{\mathbf{G}}_2&=\mathbf{U}_{g,2}\boldsymbol{\Lambda}^{(1/2)}_g\mathbf{V}^H_g,\\
      \mathbf{U}_g&=[\mathbf{U}^T_{g,1},\mathbf{U}^T_{g,2}]^T,
      \end{aligned}
      \end{equation}
      where the dimension of $\mathbf{U}_h$,$\boldsymbol{\Lambda}_h$,$\mathbf{V}_h$ are $N_r\times N_r$,$N_r\times N_r$,$2N_t\times N_r$, respectively and the dimension of $\mathbf{U}_g$,$\boldsymbol{\Lambda}_g$,$\mathbf{V}_g$ are $2N_t\times N_r$,$N_r\times N_r$,$N_r\times N_r$, respectively. In addition, SVD of $\mathbf{F}_i$ and $\mathbf{F}_r$ are given by $\mathbf{F}_i=\mathbf{X}_i\boldsymbol{\Lambda}^{(1/2)}_i\mathbf{Z}_i$ and $\mathbf{F}_r=\mathbf{X}_r\boldsymbol{\Lambda}^{(1/2)}_r\mathbf{Z}_r$.

\textit{Proposition 1:} By using equivalent decomposition, the mse relation in (23) can be maximized such that
\begin{equation}\label{eq24}
\begin{aligned}[b]
      \mathbf{X}_r&={\mathbf{V}}_g,
      \mathbf{Z}_r={\mathbf{U}}_h,\\
      \mathbf{X}_1&={\mathbf{V}}_{h,1},\\
      \mathbf{X}_2&={\mathbf{V}}_{h,2}.
      \end{aligned}
      \end{equation}

      \textit{Proof}:
Denote $\mathbf{T}$ and $\mathbf{R}$ are Hermitian and positive definite. Then, function $|\mathbf{I}_N+\mathbf{T}^{-1}\mathbf{R}|$, is maximized when $\mathbf{T}$ and $\mathbf{R}$ commute and have eigenvalues in opposite order. Two matrices $\mathbf{T}$ and $\mathbf{R}$ are commute when $\mathbf{T}\mathbf{R}=\mathbf{R}\mathbf{T}$. By using this Lemma, the mse is maximized when precoder and relay matrices have diagonal structure and follow proposed design.

      By replacing sources precoder and relay precoder structure in $mse_i$, $P_T$ and $P_i$, we obtain relations in (\ref{eq25}).
Even after the above transformation, the optimization problem is nonconvex over optimization coefficients. We apply a convex
optimization method to optimize the functions with respect to each variable and introduce an alternating
optimization algorithm to solve them. Therefore, we divide obtained problem in (\ref{eq25}) into three sub-problem and apply the proposed algorithm for each subproblem.

\begin{equation}\label{eq25}
\begin{aligned}
\underset{\boldsymbol{\Lambda}_r,\boldsymbol{\Lambda}_1,\boldsymbol{\Lambda}_2} {\text{max}}&      \sum_{i,j=1,i\neq j}^2|\mathbf{I}+\sigma^2_{x_i}\boldsymbol{\Lambda}_i\boldsymbol{\Lambda}_h\boldsymbol{\Lambda}_g\boldsymbol{\Lambda}_r
      ((\sigma^2_{x_i}\sigma^2_{g_j}\|\boldsymbol{\Lambda}^{(1/2)}_r\boldsymbol{\Lambda}^{(1/2)}_h\boldsymbol{\Lambda}^{(1/2)}_i\|^2
      +\sigma^2_{n_r}\sigma^2_{g_j}\|\boldsymbol{\Lambda}^{(1/2)}_r\|^2+\sigma^2_{n_j})\mathbf{I}+\sigma^2_{n_j}\boldsymbol{\Lambda}_r\boldsymbol{\Lambda}_g)^{-1}|\\
      \text{s.t.}~~~
      &\sigma^2_{x_1}tr(\boldsymbol{\Lambda}_r\boldsymbol{\Lambda}_1\boldsymbol{\Lambda}_h)+\sigma^2_{x_2}tr(\boldsymbol{\Lambda}_r\boldsymbol{\Lambda}_2\boldsymbol{\Lambda}_h)+\sigma^2_{n_r}tr(\boldsymbol{\Lambda}_r)\leq P_{r,t}\\
      &\sigma^2_{x_i}tr(\boldsymbol{\Lambda}_i)\leq P_{i,t},i=1,2
\end{aligned}
\end{equation}

Firstly, by introducing auxiliary variables $t_k$ and $t'_k$, (\ref{eq25}) is transformed to relation in (\ref{eq26}). For each subproblem, we introduce slack variables $\beta_k,\beta'_k$ as a upper bound for the denominator of relation in (\ref{eq26}) and define $f(t_k,\beta_k)=\beta_k(t_k-1)$ and $f(t'_k,\beta'_k)=\beta'_k(t'_k-1)$. To deal with nonconvex constraints $f(t_k,\beta_k)$ and $f(t'_k,\beta'_k)$, we replace them by its convex upper bound and iteratively solve the resulting problem by judiciously updating the variables until convergence. To this end, for a given $\phi_k$ for all $k$, we define $G(t_k,\beta_k,\phi_k) \triangleq \frac{\phi_k}{2} \beta_k^2+\frac{1}{2\phi_k}(t_k-1)^2$ which obtain by considering the inequality of arithmetic and geometric means of $\phi_k\beta_k^2$ and $\phi_k^{-1}(t_k-1)^2$ and $\phi_k=\frac{t_k-1}{\beta_k}$. This procedure is also applied for $f(t'_k,\beta'_k)$.
\begin{equation}\label{eq26}
\begin{aligned}
\underset{x,t_k,t'_k} {\text{max}}    &~~~~~~~~~~~~~~~~~~~~~~~~~~~~~~\prod_{k=1}^{N_r} t_k
      +\prod_{k=1}^{N_r}t'_k\\
            \text{s.t.}~~~
&\frac{\sigma^2_{x_1}\lambda_{1,k}{\lambda}_{h,k}{\lambda}_{g,k}\lambda_{r,k}}{
      (\sigma^2_{x_1}\sigma^2_{g_2}\|\boldsymbol{\Lambda}^{(1/2)}_r\boldsymbol{\Lambda}^{(1/2)}_h\boldsymbol{\Lambda}^{(1/2)}_1\|^2
      +\sigma^2_{n_r}\sigma^2_{g_2}\|\boldsymbol{\Lambda}^{(1/2)}_r\|^2+\sigma^2_{n_2})+\sigma^2_{n_2}{\lambda}_{r,k}\lambda_{g,k}}\geq t_k-1\\
&\frac{\sigma^2_{x_2}\lambda_{2,k}{\lambda}_{h,k}{\lambda}_{g,k}\lambda_{r,k}}{
      (\sigma^2_{x_2}\sigma^2_{g_1}\|\boldsymbol{\Lambda}^{(1/2)}_r\boldsymbol{\Lambda}^{(1/2)}_h\boldsymbol{\Lambda}^{(1/2)}_2\|^2
      +\sigma^2_{n_r}\sigma^2_{g_1}\|\boldsymbol{\Lambda}^{(1/2)}_r\|^2+\sigma^2_{n_1})+\sigma^2_{n_1}\lambda_{r,k}\lambda_{g,k}}\geq t'_k-1\\
      &\sigma^2_{x_1}tr(\boldsymbol{\Lambda}_r\boldsymbol{\Lambda}_1\boldsymbol{\Lambda}_h)+\sigma^2_{x_2}tr(\boldsymbol{\Lambda}_r\boldsymbol{\Lambda}_2\boldsymbol{\Lambda}_h)+\sigma^2_{n_r}tr(\boldsymbol{\Lambda}_r)\leq P_{r,t}\\
      &\sigma^2_{x_i}tr(\boldsymbol{\Lambda}_i)\leq P_{i,t},i=1,2
\end{aligned}
\end{equation}

By applying the mentioned procedure, it is seen that (\ref{eq26}) can be transformed to second order cone programming (SOCP) over each variable. The SOCP representation of (\ref{eq26}) is shown in (\ref{eq27}). The main ingredient in arriving at the SOCP representation is the fact that hyperbolic constraint $uv\geq z^2$ is equivalent to $||[2z \quad (u-v)]^T||\leq(u+v)$.
\begin{equation}\label{eq27}
\begin{aligned}
\underset{x,t_k,t'_k} {\text{max}} &~~~~~~~~~~~~~~~~~~~~~~~~~~~~~~\tau+\tau'\\
\text{s.t.}~~~
&\|[2v_{1,j_1} ~~~  t_{2j_1-1}-t_{2{j_1}}]^T\|\leq t_{2j_1-1}+t_{2{j_1}},j_1=1,2,...,2^{q-1}\\
&\|[2v_{m,j_1} ~~~  v_{m-1,2j_m-1}-v_{m-1,2j_m}]^T\| \leq v_{m-1,2j_m-1}+v_{m-1,2j_m},m=2,...,q,j_m=1,...,2^{q-m}\\
&\|[2\tau  ~~~v_{q-1,1}-v_{q-2,2}]^T\|\leq v_{q-1,1}+v_{q-2,2}\\
&\|[2v'_{1,j_1} ~~~  t'_{2j_1-1}-t'_{2{j_1}}]^T\|\leq t'_{2j_1-1}+t'_{2{j_1}},j_1=1,2,...,2^{q-1}\\
&\|[2v'_{m,j_1} ~~~  v'_{m-1,2j_m-1}-v'_{m-1,2j_m}]^T\| \leq v'_{m-1,2j_m-1}+v'_{m-1,2j_m},m=2,...,q,j_m=1,...,2^{q-m}\\
&\|[2\tau'  ~~~v'_{q-1,1}-v'_{q-2,2}]^T\|\leq v'_{q-1,1}+v'_{q-2,2}\\
&{\sigma^2_{x_1}\lambda_{1,k}{\lambda}_{h,k}{\lambda}_{g,k}\lambda_{r,k}} \geq \frac{\phi_k}{2} \beta_k^2+\frac{1}{2\phi_k}(t_k-1)^2\\
      &(\sigma^2_{x_1}\sigma^2_{g_2}\|\boldsymbol{\Lambda}^{(1/2)}_r\boldsymbol{\Lambda}^{(1/2)}_h\boldsymbol{\Lambda}^{(1/2)}_1\|^2
      +\sigma^2_{n_r}\sigma^2_{g_2}\|\boldsymbol{\Lambda}^{(1/2)}_r\|^2+\sigma^2_{n_2})+\sigma^2_{n_2}\lambda_{r,k}\lambda_{g,k}\leq \beta_k\\
&{\sigma^2_{x_2}\lambda_{2,k}{\lambda}_{h,k}{\lambda}_{g,k}\lambda_{r,k}}\geq \frac{\phi_k}{2} {\beta'}_k^2+\frac{1}{2\phi_k}(t'_k-1)^2\\
&
      (\sigma^2_{x_2}\sigma^2_{g_1}\|\boldsymbol{\Lambda}^{(1/2)}_r\boldsymbol{\Lambda}^{(1/2)}_h\boldsymbol{\Lambda}^{(1/2)}_2\|^2
      +\sigma^2_{n_r}\sigma^2_{g_1}\|\boldsymbol{\Lambda}^{(1/2)}_r\|^2+\sigma^2_{n_1})+\sigma^2_{n_1}\lambda_{r,k}\lambda_{g,k}\leq {\beta'}_k\\
      &\sigma^2_{x_1}tr(\boldsymbol{\Lambda}_r\boldsymbol{\Lambda}_1\boldsymbol{\Lambda}_h)+\sigma^2_{x_2}tr(\boldsymbol{\Lambda}_r\boldsymbol{\Lambda}_2\boldsymbol{\Lambda}_h)+\sigma^2_{n_r}tr(\boldsymbol{\Lambda}_r)\leq P_{r,t}\\
      &\sigma^2_{x_i}tr(\boldsymbol{\Lambda}_i)\leq P_{i,t},i=1,2
\end{aligned}
\end{equation}

After convergence iterations and replacing obtained matrices $\mathbf{F}_i$ and $\mathbf{F}_r$ in optimization problem, we minimize MSE function over $\mathbf{C}_i$. The optimum $\mathbf{C}_i$ can be obtained by using proposed approach in \cite{R}.
\section{Simulations and  Discussion}
In this section, we present the computer simulation results of our proposed robust non-linear THP transceiver design. We simulate a MIMO two-way relay system with $N_r=N_t=4$. The channel matrices are modeled by copmlex Gaussina random variables zero mean and unit variance. Noise variances at the
relay and at the receivers are also assumed similar and equal to $\sigma^2_{n_r}=\sigma^2_{n_1}=\sigma^2_{n_2}=0.1$. All simulation results were averaged over 1000 independent realizations of the fading channels.

Fig. 2 depicts the convergence behavior of the proposed optimization algorithm and its required number of iterations
for different power constraint on the transmitters. This figure confirms that algorithm converge after a few iterations.
  \begin{figure}[t]
   \centering
\includegraphics[width=0.8\textwidth]{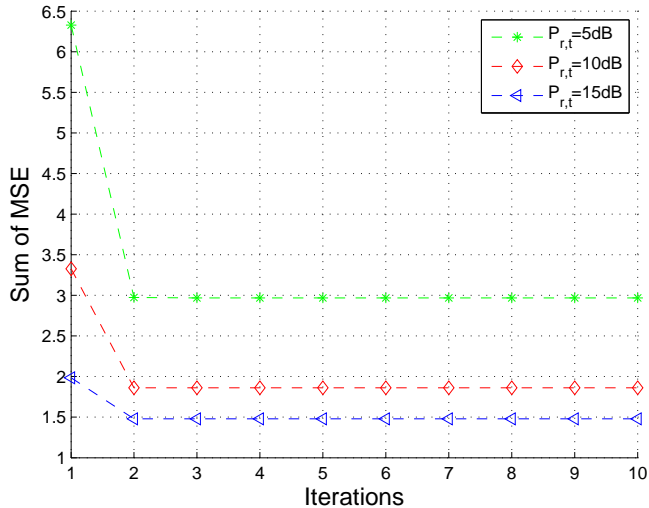}
\caption{Sum of MSE versus the number of iterations for different values of $P_{r,t}$, where $\sigma^2_{g_1}=\sigma^2_{g_2}=\sigma^2_g=0.01$.}
\label{fig1}
    \end{figure}
Fig. 3 displays the effect of channels uncertainty.
Two error bounds $\sigma^2_g = 0.01, 0.05$ are considered. The
ideal case with perfect CSI, i.e. $\sigma^2_g = 0$, is also considered. When $\sigma^2_g$ is increased, the uncertainty in channel coefficients
grows. Therefore, the MSE is increased with increasing channel uncertainty.

   \begin{figure}[t]
    \centering
{\includegraphics[width=0.8\textwidth]{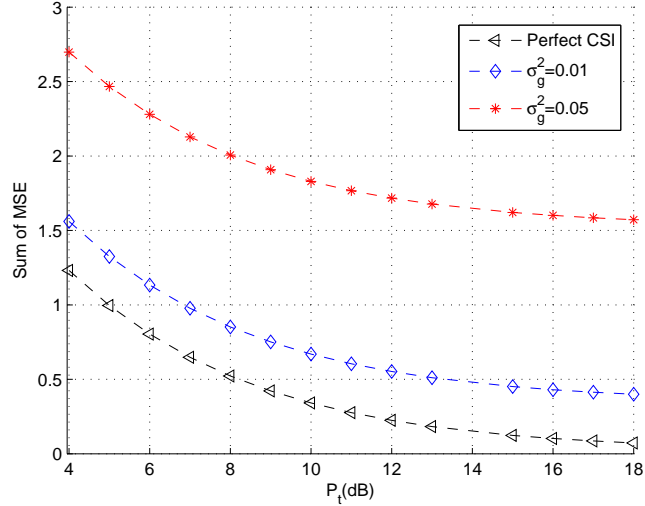}
     \caption{Sum of MSE versus the threshold of transmit power, $P_{1,t}=P_{2,t}=P_t$, for different values of channel uncertainty, where $\sigma^2_{g_1}=\sigma^2_{g_2}=\sigma^2_g$.}}
\label{fig1}
    \end{figure}
\section{Conclusion}
This paper studied the robust TH precoding for two relay network. It is assumed that the CSI is imperfect. We aim to minimize the maximum of the sum of MSE subject to transmit power of relay and transmitters is lower than a predefined threshold. The spherical model is used to characterize uncertainty of the channels. We show that the problem can be transformed to an iterative SOCP procedure. Simulations are shown to verify the efficiency of the robust algorithm.

\end{document}